# The Hubble constant from $^{56}$Co–powered Nebular Candles

*Pilar Ruiz–Lapuente*[1,2]

*Running title:* Hubble constant

[1]Max–Planck–Institut für Astrophysik, Karl–Schwarzschild–Strasse 1, D–85740 Garching, Germany. E–mail: pilar@MPA–Garching.MPG.DE

[2]Department of Astronomy, University of Barcelona, Martí i Franqués 1, E–08028 Barcelona, Spain. E–mail: pilar@mizar.am.ub.es

## ABSTRACT

Type Ia supernovae (SNe Ia), produced by the thermonuclear explosion of white dwarf (WD) stars, are used here to derive extragalactic distances and an estimate of the Hubble constant from their emission signatures at late phases (*Nebular SNe Ia Method*, NSM). The method, first developed in Ruiz–Lapuente & Lucy (1992), makes use here of an improved modeling of the forbidden line emission at late phases. Hydrodynamic models of the explosion of WDs of different masses, both sub–Chandrasekhar and Chandrasekhar, provide the basis for comparison with the observations. It is shown that it is possible to probe the overall density structure of the ejecta and the mass of the exploding WD by the effect that the electron density profile has in shaping the forbidden line emission of the iron ions, and that a robust diagnostic of the mass of the exploding WD can be obtained. Cosmic distance scale can thus be related to basic diagnostics of excitation of iron lines. Once the most adequate model is selected, comparison of the predicted line emission at these phases with the observed spectra gives an internal estimate of both the reddening and the distance to the SNe Ia. The results presented here favor denser models than those corresponding to sub–Chandrasekhar explosions. From a sample of seven SNe Ia in Leo, Virgo, Fornax and beyond, a value of the Hubble constant $H_0 = 68\ \pm 6\ (stat) \pm 7\ (syst)\ km\ s^{-1}\ Mpc^{-1}$ is derived. The depth of the Virgo cluster is found to be large, ranging from 13 to 23 Mpc at least. If NGC 4526 traces well the core of the Virgo Cluster, then the latter is located at $16 \pm 2\ Mpc$. The galaxy NGC 3267 in Leo appears to be located at 9.8 ± 1.5 Mpc.

## 1. Introduction

The Hubble constant is a key to the model of our Universe. The long–standing debate on its value has recently been rekindled by the use of the HST to observe extragalactic Cepheids (Freedman et al. 1994; Sandage et al. 1994; Tanvir et al. 1995). When comparing results from different methods to determine extragalactic distances, the use of the luminosity of Type Ia supernovae (SNe Ia) at maximum as a "standard candle" has been at odds with other widely used distance indicators (Branch & Miller 1993; Tammann & Sandage 1995). Only recently it has been clearly confirmed the correlation of brightness at maximum with rate of decline of the light curve of SNe Ia, earlier suggested by Pskovskii (1984). The use of maximum brightness of SNe Ia has to be reconsidered on the light of this correlation (Phillips 1993; Hamuy et al. 1995; Riess, Press, & Kirshner 1995). All empirical uses of SNe Ia luminosities require a zero–point calibration via Cepheid distances to a galaxy host of a SNe Ia. A different approach, which gives the distances to SNe Ia in a single step, is to use theoretical light curves (Arnett, Branch, & Wheeler 1985; Höflich & Khokhlov 1995). However, the problem there is the existing limitations in discriminating among the underlying explosion models.

Here we present also a "theoretical" approach (no need of any zero–point calibration and based on the assessment of models), but it differs from the light curve approach: to determine distances to SNe Ia we use the information contained in the emission of the SN ejecta at late phases. As in other astrophysical contexts, forbidden emission can be exploited here, since such emission of iron ions ($Fe^{+}$ and $Fe^{++}$) traces very well the electron density ($n_e$) and electron temperature ($T_e$) of the supernova "nebula". This fine sensor of density profile probes, as we will show, the mass of the exploding white dwarf. At these phases, the ejecta are kept warm by thermalization of the $\gamma$–rays and positrons ($e^{+}$) from the decay of $^{56}$Co (coming from $^{56}$Ni) into $^{56}$Fe. This is the powerhouse which sets the temperature of the ejecta. As in other "nebular" environments, the emissivity of the different forbidden transitions of $Fe^+$ keeps track of the electron density profile whereas both $Fe^+$ and $Fe^{++}$ give information on $T_e$. The large number of forbidden lines over the wavelength range at which SNe Ia are observed gives, in addition, enough information to obtain an internal estimate of the reddening. The basis of the method has been presented in Ruiz–Lapuente & Lucy (1992, Paper I), and in later work the treatment of radiation processes has been refined (see Ruiz–Lapuente 1995, R95, for a review). The input for the present study is provided by different models of the explosions, some of which involve multidimensional hydrodynamic calculations.

The explosion mechanism of SN Ia is currently being reanalyzed on basis to more consistent physical treatments than the earlier phenomenological approaches. Two classes of models for the thermonuclear explosion of WDs as SNe Ia are now competing: (1) central ignition of a C+O WD close to the Chandrasekhar mass, and (2) detonation of a C+O WD below the Chandrasekhar mass (sub–Chandrasekhar WD) induced by detonation of a helium layer accumulated on the surface. The issue is relevant to the determination of $H_0$, since as addressed in Livne & Arnett (1995), if sub–Chandrasekhar WDs do explode, the distance scale derived from SNe Ia should be revised: light curves would place SNe Ia at shorter distances. The longer distance scale is linked to the explosion of C+O WDs reaching the Chandrasekhar mass (Arnett, Branch & Wheeler 1985). An important goal of current discussions on explosion models is thus to clarify which distance scale is actually favored by SNe Ia. In the present work, by addressing the distance scale problem through modeling of the late phases of SNe Ia, we equally obtain insight on the masses of the exploding WDs. In this *Letter* we apply the *Nebular SNe Ia Method* (NSM) —or $^{56}$*Co–Powered Nebular Candle Method* (more briefly *Nebular Candle Method*, NCM), as the reader might prefer—, to a sample of SNe Ia, well observed in their nebular phases. These SNe Ia are in galaxies of the Virgo Cluster, Fornax, and Leo Group. Our method has the advantage that it does not depend on any zero–point calibration of the luminosities, and it provides individual distances to each SN Ia without assuming them to be "standard candles". It uses the "fine sensitivity" of nebular diagnostics for ejecta which, at the phases considered (about 8–10 months after explosion) do not show any evidence of clumpiness or asymmetry in their spectral lines. A major advantage of our method is the possibility of directly probing the electron density profile $n_e(r)$ and the mass of the exploding WD. This possibility arises from the fact that the electron density at the center of the ejecta $n_e(0)$ is directly linked to the mass of the exploding WD, and the forbidden emission of the Fe ions is a tracer of the density profile of the ejecta down to the center.

## 2. Physical model

The basic idea of our approach, presented in Paper I, has later been developed with the use of realistic density profiles for the underlying explosions and by significantly improving the physical treatment of the radiation transport. In the present calculations we include a numerical treatment of the escape of positrons, which takes into account their energy spectrum and follows their path throughout the ejecta for different configurations of the magnetic field: a turbulent magnetic field and a radially combed out one. In this way we estimate the amount of positron energy deposited in the different layers of the ejecta. A detailed treatment of $\gamma$–ray transport is also done, but positron deposition of kinetic energy is dominant at the times considered. Nonthermal processes resulting from the various interactions of the $\gamma$–ray photons and positrons from $^{56}$Co decay with the electrons and ions of the supernova plasma enter in the ionization balance. The ionization state is calculated with due account of photoionization arising in the recombination of the different ions. The rate equations are solved for forty most abundant ions. The electron temperature is self–consistently calculated from the balance between cooling by forbidden line emission and heating by deposition of energy by $\gamma$-rays and positrons (Ruiz–Lapuente, Kirshner, Phillips, et al. 1995, RKP95). We include all fine structure transitions up to the term $a^2I$ for FeII and up to the term $^3D$ for Fe III, for which accurate collisional data from the Opacity Project exist (Pradhan & Berrington 1993; Bautista & Pradhan 1995). A full description of the physics can be found elsewhere (R95), and a complete discussion will be done in subsequent papers. *The key diagnostics on density, however, are beyond the uncertainties on either atomic data or ion models. A basic "form–factor" discriminating between models is linked to excitation, where uncertainties of atomic data are well delimited.* For the study of the density diagnostics, rate–equation decoupled from the rest of the treatment is used for complex $Fe^+$ and $Fe^{++}$ models (54 fine–structure energy levels are included for $Fe^+$ and 34 for $Fe^{++}$) to study the variation of emissivity with electron density.

The outcome of central ignition of C+O WDs with masses close to the Chandrasekhar mass has been calculated with one-dimensional hydrodynamic codes. The resulting models, such as the popular W7 (Nomoto, Thielemann, & Yokoi 1984), or DD4 (Woosley & Weaver 1995), although lacking a consistent treatment of thermonuclear flame instabilities and propagation, have achieved reasonable success in their comparison with observations (Branch et al. 1985; Eastman & Pinto 1993; RKP95). In the alternative mechanism for SNe Ia, detonation of sub–Chandrasekhar C+O WDs from compression by helium detonation in an outer shell, WDs over a broad mass range would explode (masses from 0.6 to 1.2 $M_{\odot}$). This has been explored by different authors (Livne & Glasner 1991; Ruiz–Lapuente et al. 1993; Woosley & Weaver 1994; Nomoto 1995). Two–dimensional calculations coupled with nuclear reaction networks have recently been completed by Livne & Arnett (1995). The corresponding spectra are investigated in the present work (see also R95, and work in preparation with E. Livne). The set of models investigated here represent the whole class of so–far discussed mechanisms and progenitor masses for SNe Ia (WD masses ranging from 0.6 to 1.38 $M_{\odot}$). The observational sample consists of well–documented SNe Ia, for which good, well–calibrated nebular spectra exist. Estimates of the errors in the flux scale are available (see, for critical observational issues Ruiz–Lapuente & Filippenko 1994).

## 3. WD mass and distance scale

*The density of the SNe Ia ejecta and* $H_0$.– Diagnostics on total mass and electron density profile can provide a crucial test for the current debate on SNe Ia models. In late phases, the population of the energy levels of the ions is out of LTE across a significant part of the ejecta, because the density of the electrons responsible for collisionally exciting the lines is lower than the critical density for the corresponding transitions. A clear diagnostic of low $n_e$ comes from the drop of the emission at $\lambda$5200 Å, $\lambda$4300 Å, and $\lambda$5000 Å. These emissions are due to the $Fe^+$ $a^4F$–$b^4P$ and $a^4F$–$a^4H$ transitions, whose lower energy terms can be significantly depopulated if $n_e$ is low. This makes the collisional excitation rates decrease significantly. In this respect, an interesting ratio is that of $a^4F$–$b^4P$ to $a^4F$–$a^4P$, with emissions at $\lambda$5262 and $\lambda$8617 respectively. As illustrated in Figure 1, this ratio becomes a diagnostic of mass. The infrared transitions provide a fine diagnostic on $n_e$ too: a suitable one is the ratio of the [Fe II]$\lambda$8617 to the [Fe II] 1.257 $\mu$m line. We restrict ourselves to the optical range, however, where most of the observations are being done. The emission ratios of different transitions of the same ion provide a robust test that reaches well beyond uncertainties in ionization balance, and give an approach to $n_e$ which is independent from the other probes ($n_e(0)$ is basically a function of the mass: the number density at the center, n(0), varies by a factor larger than 10 between a 0.6 and a 1.38 $M_\odot$ WD whereas the electron fraction $X_e$ is 1.5–2 along the luminous SN ejecta). In addition to the diagnostic on $n_e$, the emission of $Fe^+$ also bears interesting diagnostics of electron temperature (the ratio of the [Fe II] $\lambda$7157, in the R–band, to those in the B–band, [Fe II] $\lambda$4416, or in the I–band, such as the [Fe II] $\lambda$8617 emission).

It is found in our analysis that, at the current stage of development of models, the sub–Chandrasekhar ones appear to fall outside the region of the ($n_e$, $T_e$) plane favored by the observations of "normal" SNe Ia. That produces discrepancies between the predicted relative luminosities of the lines and those displayed by the SN. Figure 2 (top panel) shows the calculated spectrum for W7 at 300 days after explosion, compared with the spectrum of SN 1994D, a normal SN Ia. A comparison of the same SN with the sub–Chandrasekhar model showing the closest resemblance to the observations (explosion of a 0.8 $M_\odot$ C+O WD after accretion of 0.17 $M_\odot$ of He) is done in the bottom panel. The picture where the emission of some $Fe^+$ transitions drops to zero, in SNe Ia from low–mass WDs, is at odds with the observations (Figure 2, bottom panel). Such a test allows us to conclude on the model density (massive or less massive WD) quite independently from the accuracy in the ionization balance. Our generic concern, valid for the whole range of He–induced detonation models, comes from the effects that the lower densities and, in many cases, the smaller $^{56}$Ni mass, produce in the forbidden emission at late phases. The factor by which the distance scale would be shifted by adopting the best sub–Chandrasekhar models instead of the Chandrasekhar ones is 1.1. The Hubble constant would thus be centered at 75 km $s^{-1}$ $Mpc^{-1}$, instead of being centered at 68 km $s^{-1}$ $Mpc^{-1}$ as for the Chandrasekhar model. Such a shift, however, is disfavored by "normal" SNe Ia. In addition, the variety of models that He–induced detonations naturally predict does not seem to have a clear counterpart in the observations: explosions producing 0.3 $M_\odot$ of $^{56}$Ni in a low–density WD, or those with 1.1$M_\odot$ of $^{56}$Ni moving at high speeds have not yet been identified. Binary evolution effects might, however, play a role in selecting a particular, narrow range of WD masses for exploding by He–detonations (Canal, Ruiz–Lapuente, & Burkert 1996, CRB96).

From the better agreement of "normal" SNe Ia with Chandrasekhar models, we obtain distances to our sample of SNe Ia, together with internal estimates of reddening. Table 1 gives the distances and reddenings derived for the different SNe Ia, and the absolute magnitude at maximum $M^B_{max}$ deduced from these estimates. The value of the Hubble constant obtained by this method and from this sample is $H_0 = 68 \pm 6(stat)\ km\ s^{-1}\ Mpc^{-1}$ in concordance with the $M^B_{max}$ implied for the SNe Ia. The error bar quoted here includes the effect of errors in absolute flux calibration and wavelength–dependent fluxes in the observations, and uncertainties in the internal estimate of reddening. In addition, we account for a 10% systematic error in $H_0$ coming from the method (i.e. modeling uncertainties attached to photoionization and determination of the electron temperature). Our estimate of the Hubble constant results in $H_0 = 68 \pm 6(stat) \pm 7(syst)\ km\ s^{-1}\ Mpc^{-1}$. $\langle M^B_{max} \rangle$ for "normal" SNe Ia is –19.2 ± 0.3 (SN 1991T is about 0.3 mag brighter: $M^B_{max}$= –19.53 ± 0.22).

*Extragalactic distances and SNe Ia.–* The Nebular SNe Ia Method provides a new way to determine distances connected to our understanding of SNe Ia. The extragalactic distance scale given by this method, at its present stage, already indicates the considerable depth of the Virgo cluster, which should extend at least from $\mu$ = 30.6 to 31.8 (13 to 23 Mpc), since our subsample of SNe Ia located in Virgo covers such a range of distances. Fornax is, according to our result, at a distance equivalent to that of the W Cloud, in the background of the Virgo cluster. This distance is likely to be overestimated in Table 1: SN 1992A might be slightly less dense than what is predicted by the Chandrasekhar models considered here. *Which is the distance to the Virgo core?* Tammann & Sandage (1995) argue that NGC 4526 being an early–type galaxy, it is likely to trace well the distance to the core of the Virgo cluster. Given our estimate of the distance to NGC 4526, *the distance to the core would be of the order of 16 ± 2 Mpc.*

SNe Ia have been for a long time the stronghold of the *long* extragalactic distance scale. However, when using the expression "long" or "short" for astronomical distances, one should keep in mind the depth of the Virgo Cluster, and the uncertainties arising when a Hubble flow velocity has to be assigned to the above galaxies. This uncertainty is inherent to any determination. Given this situation, it seems necessary to compare distances and absolute magnitudes for SNe Ia implied by different methods.

*How does the mass of the WD relate to the value of the Hubble constant?* Values lower than 50 km $s^{-1}$ $Mpc^{-1}$ are excluded by the Chandrasekhar limit for the explosion of a C+O WD. The ejecta can not trap more energy, and the highest electron density attainable to populate the forbidden transitions is also limited by the Chandrasekhar mass: $n_e$ (0) can not be larger than a few times $10^6$ $cm^{-3}$ at 300 days. The luminosity and the emissivity in forbidden lines is thus constrained by this requirement. From observations, it is also clear that the proper range of electron densities is below that limit: $Fe^{+}$ forbidden transitions would appear too strong for a density higher than $10^6$ cm $s^{-3}$ at 300 days. Thus, $H_0 \lesssim$ *50 km* $s^{-1}Mpc^{-1}$ *is only possible if SNe Ia would come from objects more massive than the Chandrasekhar mass. An upper limit to* $H_0$ *is equally set by the disagreement of explosions of light WDs with the bulk of SNe Ia observed* (exception to this disagreement is only found in some very subluminous SNe Ia).

The link of empirical relationships with the underlying explosion mechanism can be provided by approaches which, as the one presented here, involve density diagnostics: the late emission of SNe Ia "sees" through the ejecta at phases where spherical symmetry is still preserved. It is a good tracer of density profile and of central density of the ejecta, which can be plotted into WD mass (the central density/mass relationship is univocal for WDs). We thus expect that both the debate on the two alternative scenarios, Chandrasekhar or sub–Chandrasekhar, and the cosmic distance scale can be illuminated by these $^{56}$*Co–powered nebular candles.*

Research on late emission of supernovae was started in collaboration with L.B.Lucy, whose advice on this and related subjects is highly appreciated. A large number of hydrodynamic models were explored thanks to the stimulating interest shown by Dave

Arnett, Eli Livne, Ken Nomoto, and Stan Woosley. I would like to thank as well Wolfgang Hillebrandt for his views on the explosion mechanism of SNe Ia. Access to archived observational data has been possible thanks to Alex Filippenko, Robert Kirshner, Mark Phillips, their collaborators and to the La Palma archive. The author acknowledges the allocation of observing time at the WHT (La Palma) for the project "Nebular SNe Ia and $H_0$", and financial support for this work by the Spanish DGYCIT.

## REFERENCES

Arnett, W.D., Branch, D., & Wheeler, J. C. 1985, Nature, 314, 337

Bautista, M. A & Pradhan, A.K. 1995 (preprint).

Branch, D., Doggett, J.B., Nomoto, K. & Thielemann, F.-K. 1985, ApJ, 348, 647

Branch, D., & Miller, D.L. 1993, ApJ, 405, L5

Canal, R., Ruiz–Lapuente, P., & Burkert. A. 1996, ApJ, 456, L101 (CRB96)

Eastman, R. G. & Pinto, P. A. 1993, ApJ, 412, 713

Freedman, W.L., et al. 1994, Nature, 371, 757

Hamuy, M., et al. 1995, AJ, 109, 1

Höflich, P. & Khokhlov, A. 1995, ApJ, in press

Livne, E., & Glasner, A. S. 1991, ApJ, 370, 272

Livne, E., & Arnett, D. 1995, ApJ, 452, 62

Nomoto, K. 1995, private communication

Nomoto, K., Thielemann, F.–K., and Yokoi, K. 1984, ApJ, 286, 644

Phillips, M.M. 1993, ApJ, 413, L105

Pradhan, A.K., & Berrington, K.A. 1993, J. Phys. B, 26, 157

Pskovskii, Y. 1984, Soviet Astron., 28, 658

Riess, A. G., Press, W.H., Kirshner, R.P. 1995, ApJ 438, L17

Ruiz–Lapuente, P. 1995, in Thermonuclear Supernovae, ed. R. Canal & P. Ruiz–Lapuente (Dordrecht: Kluwer), in press

Ruiz–Lapuente, P., & Filippenko, A. V. 1994, in IAU Colloq. 145, Supernovae and Supernova Remnants, ed. R. McCray (Cambridge: Cambridge Univ. Press), in press

Ruiz–Lapuente, P., & Lucy, L. B. 1992, ApJ, 400, 127 (Paper I)

Ruiz–Lapuente, P., et al. 1993, Nature, 365, 728

Ruiz–Lapuente, P., Kirshner, R.P., Phillips, M.M., Challis, P.M., Schmidt, B.P., Filippenko, A.V., & Wheeler, J.C. 1995, ApJ, 439, 60 (RKP95)

Sandage, A., Saha, A., Tammann, G.A., Labhardt, L., Schwengeler, H., Panagia, N., & Macchetto, F.D. 1994, ApJ 423, L13

Tammann, G.A., & Sandage, A. 1995, ApJ, 452, 16

Tanvir, N. R., Shanks, T., Ferguson, H. C., & Robinson, D. R. T. 1995, Nature, 377, 27

Woosley, S.E., & Weaver, T.A. 1994, ApJ, 423, 371

Woosley, S.E., & Weaver, T.A. 1995, in Proc. Les Houches, Session LIV, Supernovae, ed. S.A. Bludman, R. Mochkovitch & J. Zinn–Justin (Amsterdam: Elsevier), 63

---

Table 1: **Distances from Nebular SNe Ia**

| SN | Galaxy | Type | (m–M)$_0$ d (Mpc) | E(B-V)[a] | $M^B_{max}$ | Group / Cluster |
|---|---|---|---|---|---|---|
| SN 1989M | NGC 4579 | Sb | 31.61± 0.2 | 0.0– 0.1 | –19.11± 0.25 | Virgo |
| | | | 21 ± 2 | | | |
| SN 1990N | NGC 4639 | Sb | 31.81± 0.2 | 0.0– 0.1 | –19.31± 0.25 | Virgo |
| | | | 23 ± 2 | | | |
| SN 1992A[b] | NGC 1380 | S0 | $31.9^{+0.17}_{-0.40}$ | 0.0–0.1 | $-19.54^{+0.45}_{-0.22}$ | Fornax |
| | | | $24^{+2}_{-4}$ | | | |
| SN 1991T | NGC 4527 | Sb | 30.57 ± 0.17 | 0.1–0.15 | –19.53± 0.22 | Virgo |
| | | | 13 ±1 | | | |
| SN 1991aa[c] | An1242-06 | SB | 33.12 ± 0.16 | 0.0–0.1 | ...[d] | $v_0$=3300 [e] |
| | | | 42 ± 3 | | | |
| SN 1989B | NGC 3627 | SB | 29.95 ± 0.4 | 0.3 | –18.81 ± 0.45 | Leo Group |
| | | | 9.8 ± 1.5 | | | |
| SN 1994D | NGC 4526 | S0 | 31.02 ± 0.25 | 0.0–0.1 | –19.12 ± 0.3 | Virgo |
| | | | 16±2 | | | |

[a] Internal estimate of reddening

[b] This SN Ia might be below 1.38 $M_\odot$

[c] The coordinates of the galaxy are: $\alpha = 12^h\ 42^m\ 34.7^s$, $\delta$=-06° 02′ 38.3″ (eq. 1950.0)

[d] unknown $m_B$ at maximum

[e] $v_0$: observed recession velocity of the parent galaxy in km s$^{-1}$

## Figure Captions

Figure 1 The electron density profile of Chandrasekhar and sub–Chandrasekhar SNe Ia ejecta and its effect in producing different ratios of $Fe^{+}$ forbidden emission. The [Fe II] $\lambda5262/\lambda8617$ ratio is displayed as a function of $n_e$, for the maximum and minimum electron temperatures reached in those models (short dashed lines). For each model, the maximum electron density ($n_e$) at each phase, here 270 (1) and 300 (2) days after explosion) is given by the intersection of the $n_e$ profile (lines running from upper right to lower left) with the x-axis, and the projection of this value on the curve of the emissivity ratio gives an estimate of the relative intensities of the lines for that model.

Figure 2 Comparison of spectra of SNe Ia (observed spectrum: solid lines; model spectra: dotted lines). The spectrum of SN 1994D in NGC 4526 is plotted together with that of the Chandrasekhar W7 model (Nomoto, Thielemann, & Yokoi 1984) at 300 days after explosion (top panel). A similar comparison is done (bottom panel) with the sub–Chandrasekhar 2D Model 6 from Livne & Arnett (1995), which yields an amount of $^{56}$Ni equal to that of model W7 but in the explosion of a 0.8 $M_\odot$ WD after accretion of 0.17 $M_\odot$ of He. Arrows indicate evidence of a too low–density model from ratios of forbidden line emission of $Fe^{+}$. Stable Ni is deficient in sub–Chandrasekhar models, and it is, however, seen in “normal” SNe Ia. Its emission as [Ni II] $\lambda7379$ is well accounted for by Chandrasekhar models. Among the class of sub–Chandrasekhar models, model 6 is the one most closely resembling “normal” SNe Ia. It is representative of He–detonations of WDs of total mass before explosion $M_{final} \approx 1 M_\odot$ which synthesize about 0.6 $M_\odot$ of $^{56}$Ni (equivalent models by other authors have similar features). Both less and more massive WDs exploding by He–detonations seem worse candidates for “normal” SNe Ia.

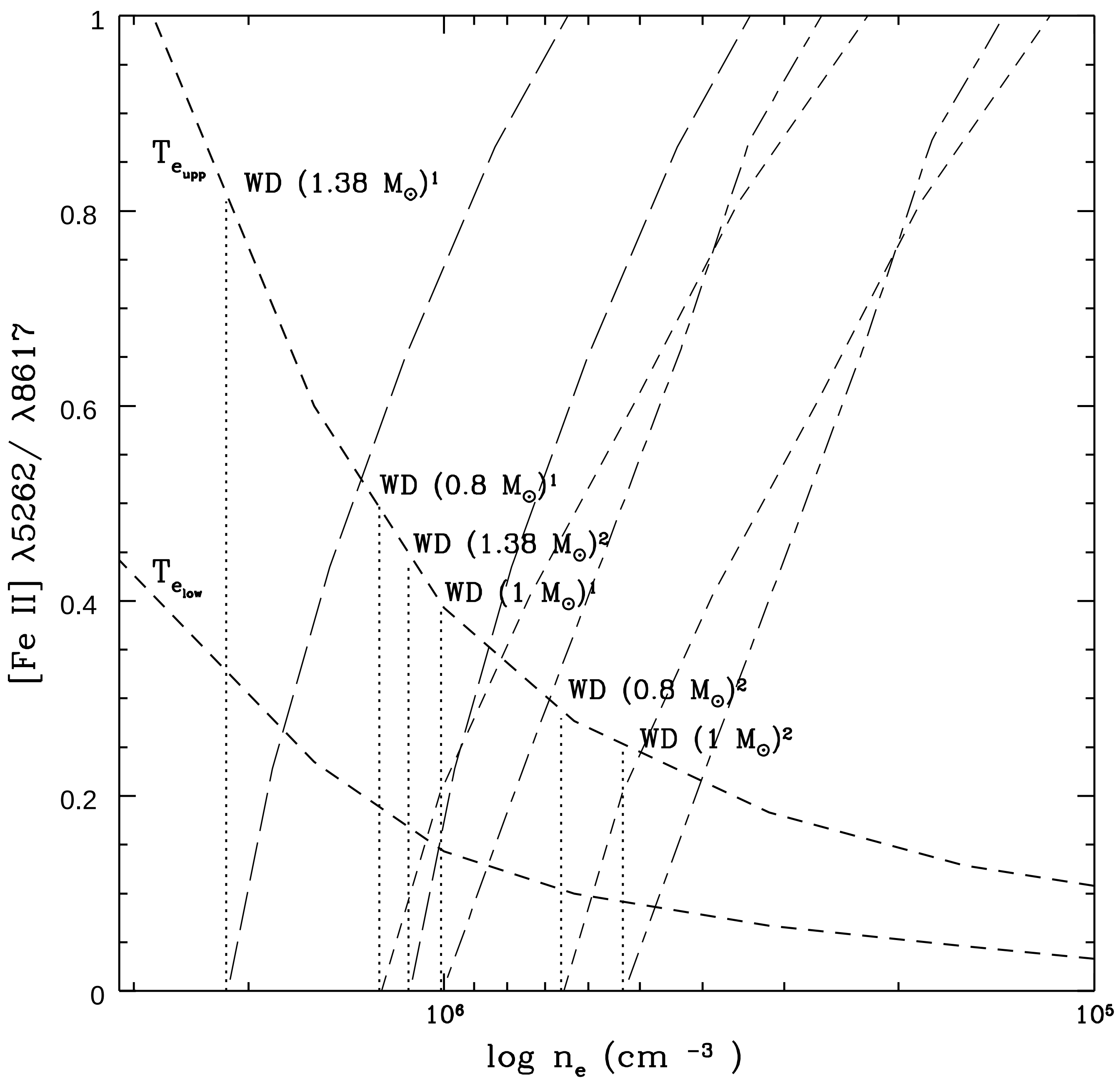

Fig. 1.—

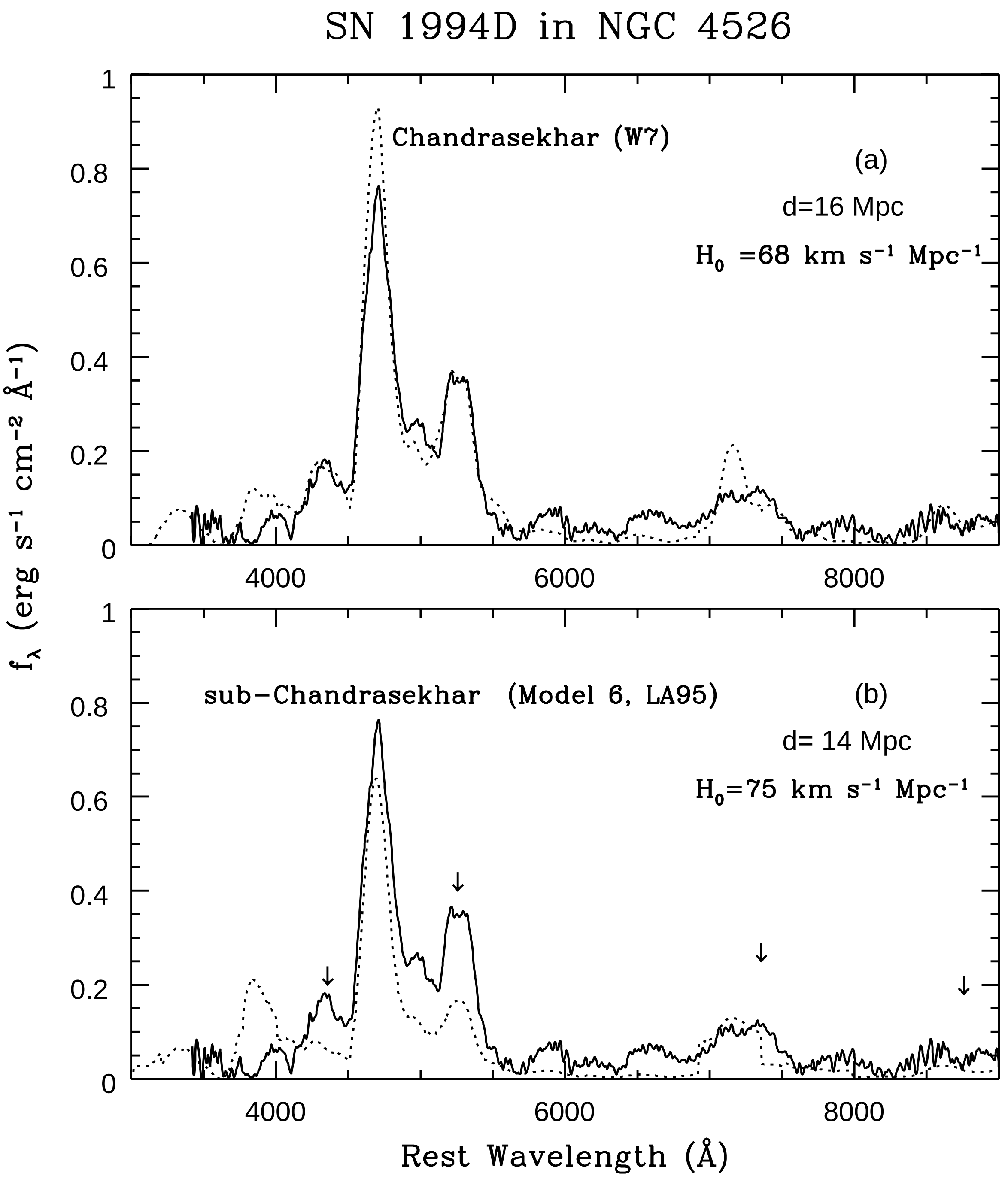
SN 1994D in NGC 4526
Chandrasekhar (W7)
(a)
d=16 Mpc
H0 =68 km s-1 Mpc-1
sub-Chandrasekhar (Model 6, LA95)
(b)
d= 14 Mpc
H0=75 km s-1 Mpc-1
fλ (erg s-1 cm-2 Å-1)
Rest Wavelength (Å)
0
0.2
0.4
0.6
0.8
1
4000
6000
8000

Fig. 2.—